\documentstyle[prl,floats,epsfig,aps]{revtex}
\begin{document}
\title{%
Theoretical models for single-molecule DNA and RNA experiments: \\
from elasticity to unzipping
}
\author{%
S. Cocco $^{\text{a}}$, J.F. Marko $^{\text{b}}$, R. Monasson $^{\text{c}}$
}
\address{%
\begin{itemize}\labelsep=2mm\leftskip=-5mm
\item[$^{\text{a}}$]
 CNRS--Laboratoire de Dynamique
des Fluides Complexes, 3 rue de l'Universit{\'e}, 67000 Strasbourg, France;
\item[$^{\text{b}}$]
Department of Physics, The University of Illinois at Chicago, 
845 W. Taylor St., Chicago, IL 60607;
\item[$^{\text{c}}$]
CNRS--Laboratoire de Physique Th{\'e}orique de l'ENS, 24 rue
Lhomond, 75005 Paris, France.
\end{itemize}
}
\bibliographystyle{unsrt} 
\maketitle
\thispagestyle{empty}
\begin{abstract}
We review statistical-mechanical theories
of single-molecule micromanipulation experiments on nucleic acids.
First, models for describing polymer elasticity are
introduced.  We then review how these models are used to interpret
single-molecule force-extension experiments on single-stranded
and double-stranded DNA.  Depending on the force and the molecules
used, both smooth elastic behaviors and abrupt structural transitions
are observed.  Third, we show how combining the elasticity 
of two single nucleic acid strands with a description of the base-pairing
interactions between them explains much of the phenomenology and kinetics of
RNA and DNA `unzipping' experiments.
\end{abstract}

\par\medskip\centerline{\rule{2cm}{0.2mm}}\medskip
\setcounter{section}{0}


\section{Introduction}

Single-molecule studies that provide information on properties of one 
or a few interacting biomolecules are becoming increasingly 
important in biophysics. The precision of control and quantitative
measurement, and simple interpretation of these experiments,
make detailed theoretical analyses appropriate.  
The theory of single molecule micromanipulation
experiments is a new development of polymer physics, emphasizing the
structural richness of biopolymers (inhomogeneity of sequence,
sequence-specific monomer interactions, transformations of
secondary structure...).   Both equilibrium and non-equilibrium
aspects of single-molecule experiments reveal new basic physical problems.

This review presents some of the theoretical ideas that have been
useful for description of single-molecule micromanipulation studies of
nucleic acids.
First, models useful for describing biopolymer elasticity will be 
presented.  We will then review how these models are used to interpret
single-molecule DNA force-extension experiments, which show
both smooth elastic behaviors and abrupt structural transitions. 
Third, we will show how combining the elasticity 
of two single nucleic acid strands with a description of the base-pairing
interactions between them explains much of the phenomenology of
RNA and DNA `unzipping' experiments.

The theoretical studies that we review use 
a wide range of tools and concepts from statistical mechanics and 
quantum mechanics. Single molecules are  
composed of a large number of elementary units (monomers).
The nearest-neighbour character of the interactions between monomers
often leads to partition functions with the form of path integrals 
(the curvilinear coordinate plays the role of time) which can be 
analyzed using the tools of quantum mechanics.
Ideas from the theory of phase transitions are also extensively employed,
for example to describe the abrupt, first-order-like  structural
changes frequently observed in stretching experiments.  The kinetics
of such transitions are thus related to problems 
from non-equilibrium statistical mechanics.
    
\section{Theoretical models of flexible polymers}

A number of polymer models have been used to model single-molecule
experiments.  Here we focus on applications relevant to 
double-stranded DNA, which is important biologically and also
nearly ideal as an object for theoretical study.

\subsection{Gaussian Model}

The simplest description of a polymer is the 
Gaussian polymer (GP) model\cite{Flo89,Doi86}, which essentially considers a 
polymer to be a series of particles joined by Hookean springs (Fig.~1). 
Let us call ${\bf r}_n=(x_n,y_n,z_n)$ 
the location of monomer $n$.
The vector leading from monomer $n-1$ to monomer $n$, 
${\bf r}_n-{\bf r}_{n-1}$,
obeys a Gaussian distribution of average zero and variance
$ < ({\bf r}_ n- {\bf r}_{n-1})^2>=b^2,$
\begin{equation}
D({\bf r}_n-{\bf r}_{n-1})= \left(\frac{3}{2\pi b^2}
\right)^{3/2}\; \exp \left ( -
\frac{3\, {(\bf r}_n-{\bf r}_{n-1}) ^2}{2 b^2} \right)
\end{equation}
where the index $n$ runs from 0 to $N$. 

\begin{figure}[t] 
\begin{center}
\psfig{figure=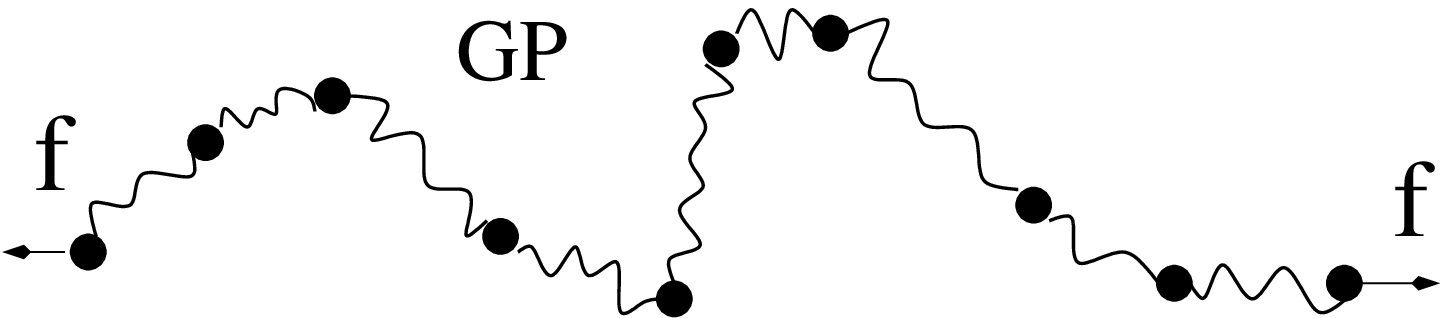,height=1.125cm,angle=0}\break
\hskip .25cm
\psfig{figure=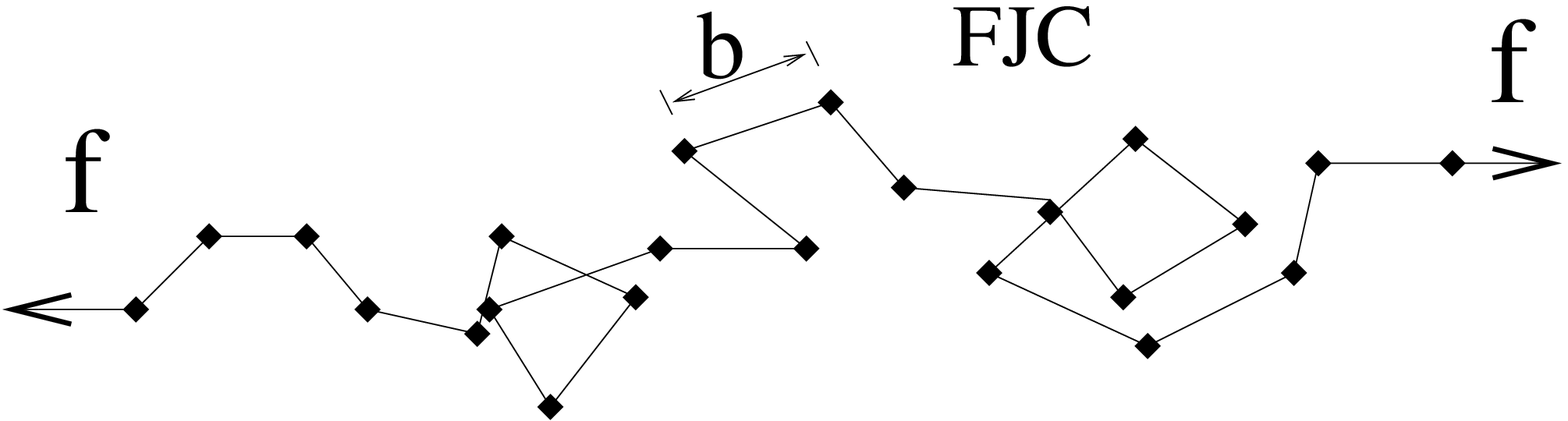,height=1.5cm,angle=0}\break
\hskip .25cm
\psfig{figure=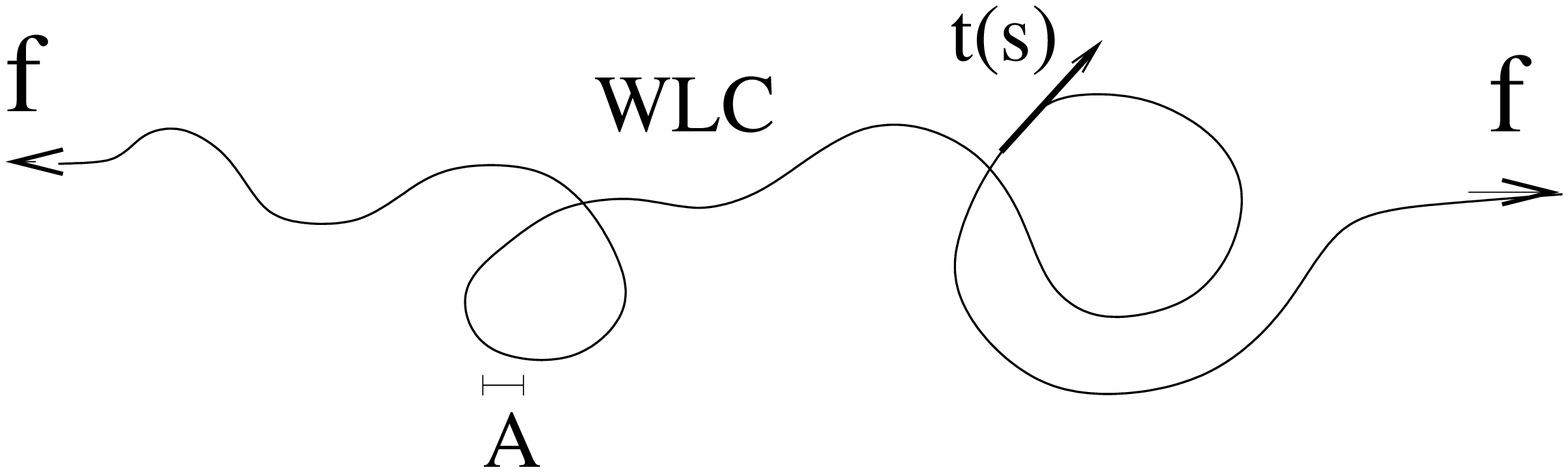,height=1.5cm,angle=0} \break
\hskip .05cm
\psfig{figure=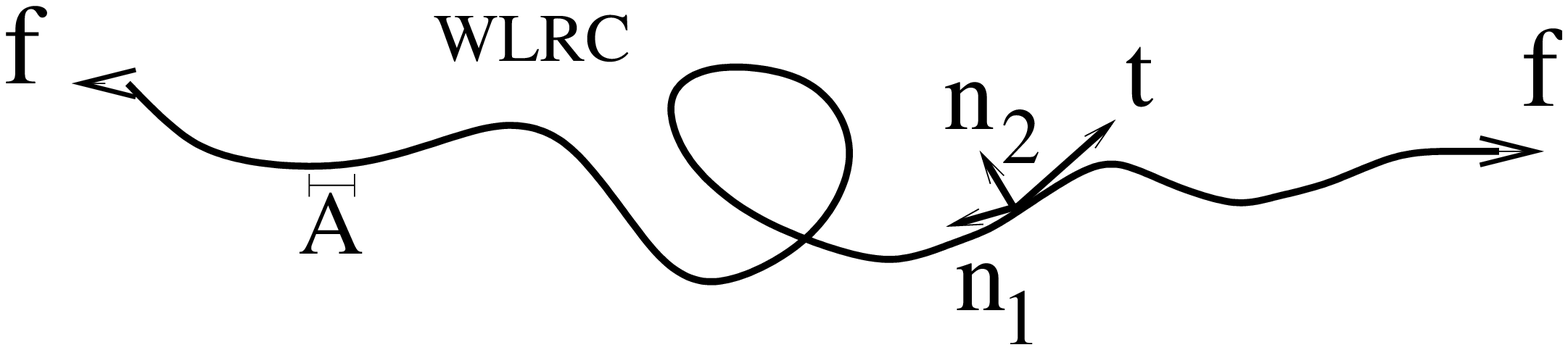,height=1.05cm,angle=0} \break
\end{center}
\caption[]{Mathematical representations of polymeric chains.
GP: the Gaussian polymer is made of $N$ monomers represented
by harmonic springs. FJC: the Freely Jointed Chain is
composed of $N$  bonds of fixed length  $b$, with no correlation
between the orientation of adjacent segments.
WLC: the Worm Like Chain is a continuous model, characterized by a
persistence length $A$; the orientation of the chain tangent ${\bf t}(s)$ is
changing appreciably over contour lengths greater than $A$. 
WLRC: the Worm Like Rod Chain is described by a rotating three-dimensional
coordinate system, with local triedron (${\bf t}, {\bf n}_1, {\bf
n}_2$), along the curvilinear coordinate $s$.  
\label{fjcfig}}
\end{figure}

When submitted to a force $f$ along
the $z$ direction, the Hamiltonian thus has a Gaussian elastic term,
and a force-distance term,
\begin{equation}
H _{GP}
=\sum_{n=0}^{N}  \frac{3\, k_B T}{2 b^2}\, ({\bf r}_n-{\bf r}_{n-1}) ^2 -  
f \,(z_N -z_0)  
\end{equation}
where the force is taken to act in the z-direction.
The statistics of the z-component of the end-to-end vector are 
easily computed; for example
the average end-to-end distance as a function of the force is
\begin{equation}
\label{spring}
<z> _{GP}= \frac{N b^ 2}{3 k_B T}\; f
\end{equation}
The Gaussian polymer as a whole behaves as a Hookean spring 
of zero rest length and stiffness $C= 3 k_B T/N b ^2$ proportional to the
temperature and inversely proportional to the length $N$. 
This effective elasticity is a model for the entropic elasticity
resulting from the decrease in the number of a polymer's configurations 
as is it extended.  This basic picture of flexible polymer
elasticity is the basis of rubber elasticity and a starting point for
polymer physics\cite{Doi86}.

\subsection{Freely-Jointed Chain model}
The GP has the unphysical feature that it can be indefinitely extended,
and is therefore useful only for weakly stretched polymers, and even
then only when physio-chemical details of the monomers are not of interest.
A model which corrects the indefinite extensibility problem but which
is still elementary and in wide use is 
the Freely Jointed Chain model\cite{Flo89,Doi86}, which
consists of $N$ bonds of fixed length $b$ (the Kuhn length), 
allowed to point in any direction independently of each other 
(see Fig~\ref{fjcfig}).  When under zero force, the mean-squared
end-to-end distance is $R^2 = N b^2$, the familiar result for
a random walk of $N$ steps.

When this chain is subjected to a force $f$, the bonds tends to align
along the force, as dipoles in an electric field, with an
energy 
\begin{equation}
\label{fjc1}
H_{FJC}=-f\,b\, \sum _{n=1}^N \cos \theta_n
\end{equation}
where $\theta_n$ is the angle between the force and the $n^{th}$ bond 
directions.  Since the segment orientations are decoupled, the partition function 
is easily calculated.  The mean average end to end distance when a
force $f$ is applied is
\begin{equation}
\label{fjc2}
<z> _{FJC}= N\, b <\cos{\theta}>= 
N\, b \left[ \coth 
\left(\frac{f\,b}{k_B\,T}\right)-\frac{k_B\,T}{f\,b}\right] 
\end{equation}
The small force behavior coincides with that of the GP expression (\ref{spring}).
However, (\ref{fjc2}) departs from the GP at large forces,
since the FJC model properly takes into account that
the extension of the molecule cannot exceed the total contour length $Nb$.
For large $f$, $<z> _{FJC}/[Nb] \approx 1 - k_B T/[b f]$.

\subsection{Worm-Like Chain}
The Worm-Like Chain (WLC)  \cite{Flo89} is a continuous
model without the sharp bends of the FJC.  The chain is described
by its unit tangent vector ${\bf{t}(s)}$, as a function of contour
length $s$ along the chain.  If no forces are applied, the tangent vector 
is presumed to undergo Gaussian fluctuations with zero mean and 
variance $<(d{\bf t}/ds)^2> = 1/A$ (Fig.~1).  The energy 
for this model in presence of a force $f$, is given by
\begin{equation}
\label{wlc}
H_{WLC}=k_BT \, \frac{A}{2}\int_0^{L} ds 
\left ( \frac{\partial {\bf t}}{\partial s} \right)^2
 -f \int_0^{L} \left ( {\bf z} . {\bf t}(s) \right) ds 
\end{equation}
The first term is curvature energy that accounts for the resistance
of the chain to bending, and the second term is the stretching energy
due to application of the external force $f$.
The partition function of a chain of length $L$, with tangent vectors 
at extremities  ${\bf t}(s=0)= { \bf t}_0$, 
${\bf t}(s=L)= {\bf t}_1$, can be written as a path integral,
\begin{equation}
\label{wlcz}
Z(L,f,{\bf t}_0,{\bf t}_1)=\int D {\bf t}\; e^{\;-{H_{WLC}}/{k_BT}}
\end{equation} 
The significance of the parameter $A$ is made clear
by the expression of  average  scalar product between 
tangent vectors at coordinates 
$s$ and $s'$ at zero force, 
\begin{equation}
<{\bf t}(s) \cdot {\bf t}(s') >=\exp\,(\,-|s-s'|/A\,)\,
\end{equation}
Therefore $A$ represents the characteristic distance above which 
tangent vectors decorrelate; $A$ is called the persistence length. 

For a long chain under zero tension the WLC mean-squared end-to-end distance
is $R^2 = 2 A L$ for $L>> A$ (the formula for general $L$ is often useful,
see \cite{Doi86}).  
Therefore the unperturbed random coil properties of the WLC are equivalent
to those of the FJC and GP if we make the identification $2A=b$ and $L=Nb$.
The unstretched WLC on large scales becomes a random walk
of $N=L/(2A)$ steps each of length $b=2A$.

From a physical point of view, the FJC represents 
$N$ uncorrelated dipoles in an electrical
field, and the average  orientation of one  dipole at equilibrium is obtained
by classical statistical mechanics (eqns \ref{fjc1}, \ref{fjc2}).
By contrast, the WLC describes, the ``time''  evolution  of a 
dipole with moment of inertia $A$ in an electric field, with the role of
time played by the contour length coordinate $s$.  The introduction 
of the time dimension makes WLC equivalent to a quantum mechanical problem. 
The Schr\"odinger equation for the associated  
wave function can be analytically solved for small and large force limits, and
can be numerically solved for general force\cite{Bus94,Mar95}.

At small forces $<< k_B T/A$ the Hookean behavior (\ref{spring}) is recovered
(i.e. with $b \rightarrow 2A$ and $N \rightarrow L/b$) while
for large forces $<z>_{WLC}/L\approx 1-\sqrt{k_B T/(4\,A\, f)}$.
\begin{figure}[t] 
\begin{center}
\psfig{figure=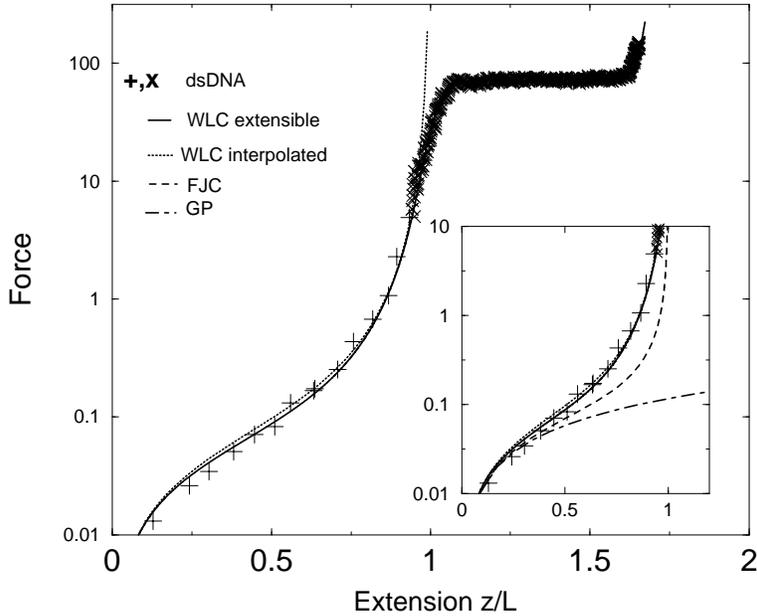,height=10cm,angle=-90}
\caption[]{Force-extension curve for a $\lambda$-dsDNA with
equilibrium contour length $L=16.4 \mu$m in  10mM $Na^+$ buffer. 
Experimental data (x) are from [24] and (+) from [14].
The plateau at $f\simeq 70$ pN indicates the cooperative
transition to SDNA. The extensible WLC model with persistence 
length $A=53$ nm, Young modulus $\gamma \approx 1000$ pN, and 
two states dsDNA-SDNA   reproduces accurately the experimental 
behavior [23]. Marko and Siggia's interpolation formula [3]
(\ref{mark2}) is very accurate up to forces of 10 pN.
Predictions from the GP and FJC ($b=2A=106$ nm) models 
are plotted in the Inset.  
\label{forzaesfig}}
\end{center}
\end{figure}
\section{Elasticity of Double- and Single-stranded DNAs: 
experiments and theory}

\subsection{Double-stranded DNA (dsDNA) under tension }
\label{tensionds}

Reviews of experiments and theory on dsDNA elasticity can be
found in \cite{Aus97,Bus00,Wan97,Str00}.   
In Fig.~2 we report the force extension curve of a single dsDNA.
Experimental data obtained by \cite{Clu96,Str98}
for $\lambda$-DNA of length 48502 bp, or $L=16.4$ microns are shown,
with fits of the GP, FJC, and 
WLC models with $A=b/2= 53$~nm in 10 mM $Na^{+}$ buffer.
For dsDNA, the Kuhn length $b=106$~nm is  much larger than the
natural  base pair spacing of 3.4 $\AA$.
dsDNA is  not  naturally described by the FJC model because
consecutive bases, stacked onto each other, are not free to reorient
independently of each other. The success of the WLC shows that
dsDNA behaves as a semiflexible polymer, with a bending modulus $A\; k_BT$.

Several analytical interpolation formulae for the WLC, and modifications
of the FJC introduced to fit accurately the data are discussed and compared
in  \cite{Wan97,Bou99}.
Marko and Siggia proposed a simple interpolation formula that
is close to the exact numerical solution of the force-extension response
of the WLC\cite{Mar95,Bus94}, 
\begin{equation}\label{mark2}
f _{WLC} (z)\simeq \frac{k_B T}{ A}\left[\frac{1}{4\,(1-z/L)^2}-\frac
14+\frac{z}{L}\right]
\end{equation}
This expression reduces to the exact solution as either $z \to 0$
or $z \to L$, but differs from the exact solution by up to $\approx 10 \%$
near $f=0.1$ pN (Fig~2). Bouchiat {\em et al.} \cite{Bou99} have introduced 
correction terms to eqn (\ref{mark2}), in the form of
a seventh order polynomial in $z/L$. The resulting approximation for
$f _{WLC}(z)$  is accurate to 0.1\%. 
According to formula (\ref{spring}), a force $f_0 = 3 k_B T/b$ 
is required to extend dsDNA by a fraction of its contour length;  
from $b\simeq 100$ nm we see that the characteristic force associated
with the entropic elasticity of dsDNA is $f_0 \approx 0.1$ pN.

Fig~2B shows that experimental data are well fitted by
the FJC model for forces $f<0.1$ pN, and by the WLC  
up to $f<5$ pN.
Various  experiments analyzed in terms of the WLC 
give $A= 50 \pm 5 $ nm in 10 mM
$Na^{+}$ \cite{Smi96,Wan97}. The  persistence length
of DNA is reduced in high salt concentrations by electrostatic
screening of the repulsive charge
along the  backbone; electrostatic effects have been
taken into account in the WLC model by Barrat and Joanny
through Debye-Huckel interactions \cite{Bar93,Mar95,Bau97,Wan97}.

Fitting larger-force experimental data demands the introduction
of the stretching elastic modulus of the molecule, 
$\gamma \simeq 1000 \pm 200$ pN, quantitatively
consistent with the relation between the bending modulus, $A\, k_B T$
and the Young modulus $Y=\gamma/(  \pi R^2)\simeq 300$ MPa
($R=10$ $\AA$ is the double helix radius) for an elastic rod,
 \begin{equation}
A =\frac{\pi}{4\, k_BT} Y \;R^4\equiv \frac{\gamma \, R^2} {4 k_B T}
\end{equation} 
The value of the elastic  modulus of DNA indicates that thermal 
fluctuations of the axial base pair distance $h$ of the order of  
$ <(h -<h>)^2>\simeq k_B T/(\gamma <h>) \simeq 0.14 \AA$. 
This order of magnitude is in agreement with molecular
dynamics simulations, providing a consistent picture of the elasticity
at the atomic and mesoscopic scale\cite{Mar03}. 
Axial vibrational  modes have been
studied in \cite{Coc99b} and compared to Raman spectroscopy measurements.

When dsDNA molecule is subjected to a force
of $f=65$ pN  it undergoes  a  structural transition to another conformation,
S-DNA, with a contour length $1.7$ times larger than its B-DNA 
counterpart\cite{Smi92,Smi96,Clu96}.
Numerical investigations of the structure of S-DNA
have been  performed by Lavery and collaborators\cite{Clu96,Leb96}.
The force plateau around $f=65$ pN corresponds to a highly cooperative
transition, reminiscent of a first order phase transition.
A two state model proposed by Cluzel
{\em et al.}\cite{Clu96,Ciz97}, inspired from models introduced in the
context of thermally-induced denaturation~\cite{Hil59,Zim59,Kafri}, 
is able to reproduce the B-to-S transition.  A recent study has
suggested that the extended S state is actually strand-separated
with the S phase described as stretched ssDNAs
\cite{Rouzina}.

In the simplest two-state model of the B to S transition,
the molecule is described as a chain of $N$ elements (base pairs), which
can be in states B (energy $E_B$, length $l_B$) or S (energy
$E_S$, length $l_S>l_B$). Each element is associated a 
spin variable, $s=1$ (respectively $-1$) for the
B (resp. S) state. The energy of the chain reads,
\begin{equation}
\label{is}
H=\sum_{i=1}^{N}(E_{s _i}-f\,l_{s _i})+
\frac{\omega}{2} \sum_{i=1}^{N-1} (1-s_i\,s_{i+1})
\end{equation}
$\omega$ represents a 'domain wall' energetic cost of a B-S
frontier. Up to an additive constant,
\begin{equation}
\label{ising}
H=-\frac{\omega}{2} \sum_{i=1}^{N-1} s_i s_{i+1} 
-\frac 1 2 (\Delta E -f \,\Delta l) \sum_{i=1}^{N} s_i 
\end{equation}
where $\Delta E=E_S-E_B$ and $\Delta  l= l_S-l_B.$
Notice that eqn (\ref{ising}) is simply the Hamiltonian of a
one-dimensional Ising model with magnetic field
$h=\Delta E -f \, \Delta l $.
The extension is obtained from the derivative of the  free energy with 
respect to the force $f$. Comparison with experiments allows
to determine quantitatively the domain wall energy, $\omega\simeq 4 k_BT$
\cite{Ciz97}.  
The extensible WLC including nonlinearities which define two states
of extension provides a way to fit the force-extension curve
over a wide range of forces $0.01<f<100$ pN \cite{Mar98}. 

\subsection{Supercoiled DNA under tension}
\begin{figure}[t] 
\begin{center}
\psfig{figure=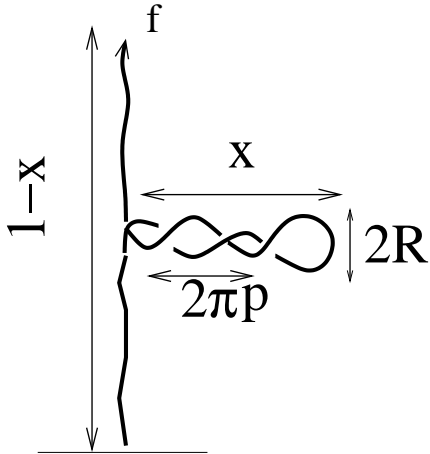,height=6cm,angle=0} 
\end{center}
\caption[]{Phase coexistence in a supercoiled DNA under tension.
A fraction $x$ of the length is plectonemic supercoil
with radius $R$ and pitch $P$, while the remaining fraction $1-x$ is in 
an extended conformation. 
\label{plecfig}}
\end{figure}

dsDNA differs from simpler polymers because it exhibits torsional 
and bending stiffness. Try to impose a twist to an elastic rod while keeping 
it extended and fixed at one end. Then, if you relieve the tension, 
an interwound structure called a plectoneme will appear, Fig~\ref{plecfig}
(twisted telephone cords often form plectonemic supercoils).
Similarly, dsDNAs under sufficient torsional stress interwinds to 
form plectonemic supercoils.  Formally, the over- or underwinding of 
DNA is measured by the change in double-helical linking number.  
This is often expressed as $\sigma$, the fractional change
in linking number relative to that of relaxed dsDNA 
(one right-handed, or positive link per 10.5 base pairs).
Supercoiling of DNA is of tremendous importance to eubacteria. 
For example, in E. coli all the DNA is held under torsional stress 
and is topologically constrained with $\sigma \approx -0.06$.
 
The elasticity of a single supercoiled DNA molecule has been
experimentally measured by Strick {\em et al.} 
\cite{Str96,Str98} and by
L\'eger {\em et al.} \cite{Leg99}. A rich behavior was observed.
At small forces the molecule responds to increasing
positive or negative supercoiling by first having its conformations
slightly chirally perturbed, and then
by forming plectonemes with appreciable shortening of its length.  At
forces $f > 0.5$ pN,  negative supercoiling is released
through the opening of the double helix into denaturation bubbles.  
At forces $f>3$ pN, positive supercoiling induces the formation
of regions exhibiting a new conformation called P-DNA.
The structure of P-DNA has been deduced by molecular modeling\cite{All98};
it is essentially characterized by its exposed bases.  P-DNA can
be thought of as two tightly interwound ssDNAs.

The theory of stretched supercoiled DNA was initiated by 
Marko and Siggia \cite{Mar95b,Mar97}, who
considered phase coexistence
of linear, plectonemic, and denatured DNA in different regions
of a supercoiled molecule. The relative extensions of these portions are
determined by the degree of supercoiling $\sigma$ and the stretching 
force $f$.
 
At small forces, a fraction $x$ of the molecule is in the 
plectonemic ($p$) regime, whereas the remaining $1-x$ fraction is 
extended ($s$) (Fig~\ref{plecfig}). The free energy per unit of length 
${\cal F}=F/L$ reads,
\begin{equation}
{\cal F} (\sigma, z/L) =x\; {\cal F}_p (\sigma_p)+ (1-x) 
\, {\cal F}_s(f,\sigma_s) 
\end{equation}
$\sigma= \Delta L_k/L^0_k$ is the excess of density of supercoiling with
respect to the rest configuration 
(dsDNA making a double helix turn in 10.4 bases,
 $L^0_k =L/10.4$);  $\sigma$ is
partitioned into extended and plectonemic regions:
$\sigma = x\, \sigma_s+ (1-x) \sigma_p$. 
The free energy of the extended phase equals
the WLC free energy plus the twisting energy,
${\cal F}_s(f,\sigma_s)=
{\cal F}_{WLC}(f)+k_BT\, C\, L_k^2 /(2 L^2)\, (2\pi\sigma_s)^ 2$. 
$C$ is the twist persistence length known from experiment to be
$C=75 \pm 30$ nm.

The free energy of the plectonemes is thus considered to be the sum of elastic 
(bending and twisting), electrostatic and entropic contributions,
minimized over plectonemic parameters e.g. the pitch $P$ and radius $R$
(Fig~\ref{plecfig}).
The entropic term represents the
entropy lost by confining the DNA in the superhelix formation. 
The total free energy is obtained by minimization with respect to the
plectonemic portion $x.$
At large forces the structural transition to denatured DNA
is also included in the model by allowing the plectonemic phase
to be a mixture of denatured and normal plectonemic DNA.
The theoretical force-extension curve at 
fixed supercoiling reproduces the experimental behavior \cite{Mar97,Str96}.

A semi-microscopic model has been used to describe thermal- 
and torque-induced denaturation in one phase diagram\cite{Coc99a}.
This work described the formation of denaturation bubbles when DNA
is stretched at $f >0.5$ pN and negatively supercoiled.
The critical torque at room temperature $\Gamma \approx -2k_BT $ is in
good agreement with the value inferred from the experiments by Strick 
{\em et al.} \cite{Str98}.

A generalization of the two-state Ising description (\ref{ising}) 
of the overstretching transition
has been introduced by Sarkar {\em et al.} \cite{Sar01} to model the
structural transition of a twisted and stretched DNA molecule
observed in \cite{Leg99}.
For each site five possible state are introduced: 
dsDNA, S-DNA, P-DNA, sc-PDNA (a supercoiled P state), and a left
handed double helix Z-DNA. This last state, with a supercoiling
degree $\sigma_Z = -1.3$, is proposed as an alternative to denatured ssDNA 
($\sigma =-1$). 
A  force-torque diagram is derived that agrees with the experiments on the
critical unwinding torque at zero force, $\simeq -2 k_B T$, and the
torque to drive DNA into the P structure, $\simeq 7k_B T.$ 

The elasticity of supercoiled DNA has also been studied at a more
microscopic level.  Twisting and bending deformations can be
described by extending the WLC to include description of base-pair
orientation using a triad of unit vectors (WLRC) (Fig~1) \cite{Fai97}.
Moroz and Nelson \cite{Mor97}, and  Bouchiat and
M\'ezard\cite{Bou98} have written the partition function of this model 
as a path integral in the space of the Euler angles
parametrizing the orientations, limiting the
integration measure to the paths with a fixed linking number $L_k$.
The free energy is obtained from the ground state energy of a Schr\"odinger
equation describing a particle moving on a unit sphere in the presence of
electric and magnetic fields.

The WLRC model does not take into account self avoidance: the WLRC chain
is a phantom chain that can pass through itself.  The result is that
linking number fluctuations are not well-defined in the continuum limit. 
This problem is reflected in a divergence of the ground state energy. 
This is strictly a technical problem since real DNA has self-avoidance
interactions.  
To avoid this problem, Moroz and Nelson considered the 
unambiguous infinite force situation (fully stretched molecule), 
and obtained finite force results by means of perturbation theory.
Bouchiat and Mezard have introduced a short distance 
cutoff (discretization of the chain) to suppress the singularity.
Fitting the theory to experimental data \cite{Str96},
the twist persistence length $C$ can be determined but is largely dependent
on the theoretical scheme followed: $C=120$ nm is obtained by Moroz and Nelson,
while $C= 82$ nm is obtained by 
Bouchiat and Mezard for a cutoff length of $7$ nm.

\subsection{Single-stranded DNA under tension}

Single-stranded DNA (ssDNA) is more flexible and can reach a larger extension
per base pair than dsDNA. 
A sensible simple model of ssDNA is, at first sight, a FJC with 
a Kuhn length equal to the sugar-phosphate monomer backbone length $b=7 \AA$.
However the ssDNA elasticity is complicated by
nucleotide interactions, and as a result simple polymer models
do not describe ssDNA elasticity over a wide range of forces.

\begin{figure}
\begin{center}
\psfig{figure=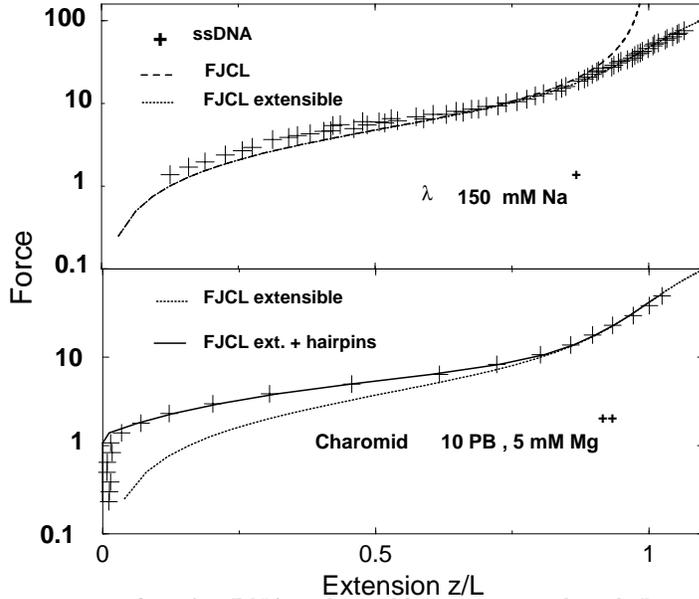,height=8cm,angle=-90} 
\end{center}
\vskip .5cm
\caption{Top: Force extension curve for a $\lambda$-ssDNA with
equilibrium contour length $L_{ss}=l_{ss} N \equiv 0.56\, nm \times
48502\,bp=27 \mu m$  in 150 mM Na$^ +$. Experimental  data (+), 
dashed line: FJCL with $b=1.5 nm$ , dotted line: extensible 
FJCL with  $\gamma=800\, pN$, from[11].
Bottom: Force extension curve for a Charomid ssDNA in 10 mM
phosphate buffer, 5 mM Mg$^{++}$ buffer, 
data are from [38], the fits are with the FJCL with $b=1.9\,
nm$ and $\gamma=800\, pN$ (dotted line) and with the hairpin model 
(full line) of [41]. }
\label{ssfext}
\end{figure}

For forces $f< 20$ pN the experimental force extension curve for a
48502 base $\lambda$ ss-DNA in 150 mM $Na^{+}$ has been fitted
 with a FJC--like 
(FJCL) model by Smith {\em et al.}  \cite{Smi96} with two effective 
parameters: a Kuhn length $d=15 \AA$~ and a contour length
per base pair  $l_{ss}=5.6 \AA$~ that differs from the backbone
distance (see Fig.~4).
Note that due to the higher flexibility, the characteristic entropic force
of the single strand, $f_0= 3 k_B T/ \sqrt{b l_{ss}} \approx 10$ pN
 (\ref{spring}), is much higher than for dsDNA. 
At forces $f>15$ pN the fit requires the introduction of a stretching modulus 
$\gamma=800$ pN (Section~\ref{tensionds}). 

ssDNA elasticity depends strongly on salt concentration.
At low salt (1 mM Na$^+$, self avoidance interactions 
due to electrostatic self-repulsion along the charged sugar-phosphate 
backbone occurs. 
The experimentally observed logarithmic-like dependence of the extension 
upon force is well reproduced by Monte Carlo simulations \cite{Des03,Zha01}. 
Electrostatic selfavoiding effects can be taken analytically into account
using Barrat and Joanny formalism \cite{Coc03}, or with
a Hartree-Fock calculation from the WLC models \cite{Bou03}.

At higher salt concentration ($>$ 100 mM Na$^+$, or in presence of Mg$^{++}$ )
formation of secondary structure (`hairpins') by hydrogen
bonding between complementary bases on the same strand strongly influences
elastic properties. Experiments show that the force--extension behavior 
curve depends on the strand GC vs. AT content, and can be modulated
using denaturing chemical agents that suppress hydrogen bonding
\cite{Mai00,Des03}.
A theoretical analysis of the elasticity of a polymer with hairpin secondary
structure has been developed by Montanari and Mezard \cite{Mon01}.
Conformations of hairpins taken into account are such that any two
pairs of paired bases ($i<j$, $k<l$) are independent
($i<j<k<l$), or nested ($i<k<l<j$) \cite{Bun99,Pag00}.
This representation do not include pseudoknots \cite{Isa00} but
 leads to a solvable 
recursion relation relating the partition functions of successively larger sequences 
\cite{Bun99}, under the simplifying hypothesis that any two bases e.g.
AT, GC, AG, ... have the same pairing free-energy.  For an infinite molecule,
a phase transition takes
place between a folded (zero extension, $f<f_s$) and an extended
phase ($f>f_s$) with $f_s$ of the order of 1 pN,  given reasonable choices
of parameters (see Fig.4).

\subsection{Elasticity of DNA in presence of DNA-folding proteins.}
In eukaryote cells, DNA is wrapped around octamers of histone
proteins to form a more compact structure called a nucleosome.
The long chromosomal DNAs of eukaryote cells 
are thus organized into long strings of nucleosomes, or
`chromatin fibers'.  A single eukaryote chromosome may contain
more than $10^8$ base pairs of DNA and roughly $10^6$ nucleosomes.

The elasticity of chromatin fiber has been experimentally studied by Cui
and Bustamante \cite{Wan98}. The experimental curves can be fitted 
with polymer models composed of units which, 
independently of each other, can be in a 
folded (short) or unfolded (long) state \cite{Mar97a,Coc03}.
These states are taken to correspond to stacked and unstacked nucleosomes.
The elastic response of whole mitotic chromosomes can be related back
to this fiber elastic response\cite{Coc03}.

Very roughly, models of DNA folding by proteins will generally show
a characteristic force at which the proteins will dissociate in
equilibrium\cite{Mar97a}.
Given a free energy difference between the folded and unfolded
states of $g$ per fold, and given an end-to-end length reduction
of $d$, this characteristic force will be about $g/d$.  Note that
for large values of $d$ (e.g. by formation of large DNA loops,
a common feature of gene-regulatory proteins) this implies
low on-off equilibrium forces.  It must be kept in
mind that if the enthalpic component of the binding free energy
is large, there may be large barriers for such loops to open
and close, making equilibrium difficult to reach.
Such situations should show theoretically interesting 
many-body thermal barrier-crossing kinetic phenomena.
 
\section{DNA and RNA unzipping}

\begin{figure}[t] 
\begin{center}
\psfig{figure=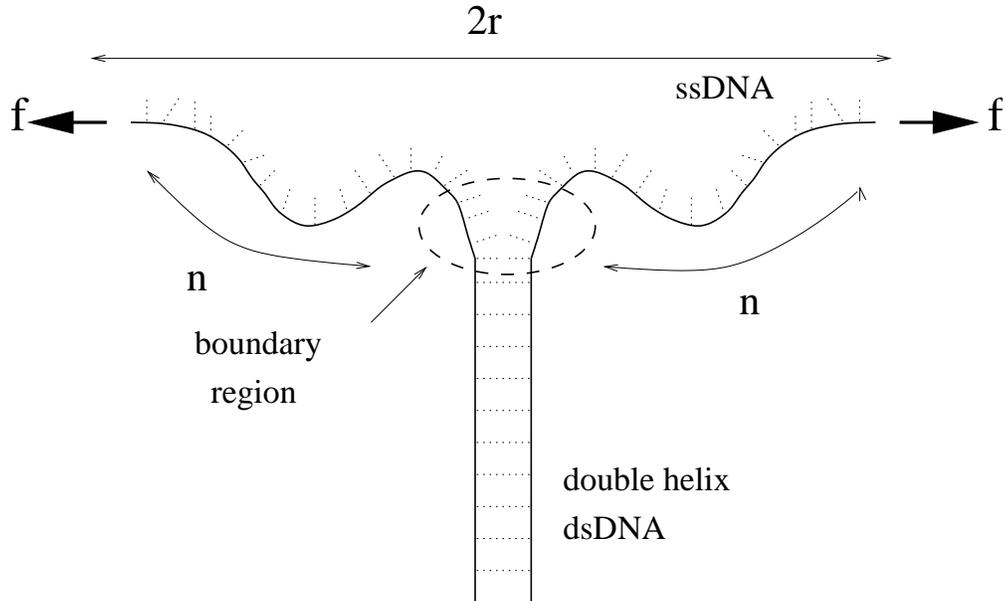,height=8cm,angle=0} 
\caption[]{Sketch of a DNA molecule with $n$ base pairs
unzipped, as a result of a mechanical stress (applied force $f$).
The distance between the two ssDNA ends is defined to be $2 r$.
\label{unzfig}}
\end{center}
\end{figure}

Essevaz-Roulet, Bockelman and Heslot
have shown that the two strands of a dsDNA
can be pulled apart by a force $\approx 12$ pN
\cite{Ess97} (Fig.~\ref{unzfig}). 
Variations of the `unzipping' force 
correspond to the DNA sequence, through the known relationship between
DNA sequence and base-pair interaction strengths\cite{Bre86}.
Higher forces were shown to correspond to DNA regions with 
higher GC densities, which in general have
stronger base-pairing interactions than AT-rich sequences.
Experimental unzipping force traces show a series of sawtooth signals 
attributed to stick-slip motion, with the sticking generated by DNA regions
with higher GC content.
This kind of experiment amounts to 'feeling' DNA sequence.

Current techniques are able to observe GC-content over long
stretches of DNA ($\approx 10$ kb) with about 10 base pair resolution.
It should be noted that it has been demonstrated that unzipping 
is sensitive to at least some single-base substitutions\cite{Boc02}.
We now discuss some equilibrium and dynamical aspects 
of DNA and RNA unzipping.

\subsection{Thermodynamics of DNA-RNA unzipping}   

Control parameters for unzipping vary from experiment to experiment. 
Roughly speaking, either the force or the distance between strand 
extremities may be kept fixed (Fig.~\ref{unzfig}).

\subsubsection{Fixed pulling force}

With a fixed  force $f$ on the 
molecule ends, the free energy $G$ of the molecule
with $n$ base pairs opened is the difference between the free energy of 
the two extended single strands of $n$ bases each, 
and the free energy lost in unpairing the $n$ first base pairs ($i=1, \ldots ,n$):
\begin{equation}
\label{unz1}
G (f,n)=2\; n\, {\cal F}_{ss} (f) -\sum_{i=1}^{n}\; g_{ds} (i)
\end{equation}
As  discussed above, the free energy per base pair of stretched ssDNA, 
${\cal F}_{ss} (f)$, can be expressed using the FJCL model for forces 
$f \simeq 10\,$ pN \cite{Smi92}. At this
high tension, nucleotide hairpin-formation effects are absent.
Quadratic expansion of ${\cal F}_{ss} (f)$ around $f=10\,$ pN 
gives the free energy of a Gaussian polymer, 
  ${\cal F}^{GP}_{ss} (f) =- f^2 b^2 /(6 k_B T) $
with an effective Kuhn length $b=7 \AA$.  

We start by considering an homogeneous sequence, where all base pairs
have pairing free energy  $g_{ds}= - g_0$.
The unzipping critical force $f_u$ is simply given by the condition 
\begin{equation}
\label{fc}
G(f,n)\equiv n\; g(f)= [2\, {\cal F}_{ss}(f)+ g_0] n = 0
\end{equation}
For $f < f_u$, dsDNA is stable, and if $f > f_u$, the double helix unzips
as in a first order phase transition.
Using the Gaussian model for the ssDNA, we obtain 
$f_u^{GP}=\sqrt{3 k_B T g_0}/b$.
The unzipping force of a homogeneous sequence is therefore directly 
related to the pairing free energy.

Rief {\it et al} measured the unzipping forces $f_u$ for DNAss of
various repeated sequences\cite{Rie99}. It was found that
$f_u$(poly~dA- dT)=9$\pm 3$~pN and $f_u$({poly~dG-dC})=20$\pm 3$~pN, giving
$g_0^{GC}({A-T})=0.8\; k_B T$,$\ g_0^{FJCL}({A-T})=1.1\; k_B T$
$g_0^{GC}({G-C})=4.2\; k_B T$, and
$g_0^{FJCL} ({G-C})=3.5\; k_B T$  respectively.
These values of the free energies of denaturation are compatible
with thermodynamic data based on DNA melting\cite{Bre86}.
It is to be noted that unzipping experiments give the only direct
measurement of the relative free energies of ss and dsDNA at equal
temperatures.  

Unzipping has been 
discussed in the language of continuous phase transitions
by Lubensky and Nelson \cite{Lub00,Lub02}. 
The unzipping free energy per base pair, $g(f)$, vanishes as $f_u- f$  
with a discontinuous slope, as in a first order phase 
transition.  However (as in interfacial wetting)
as the unzipping transition is approached from below {\em i.e.} $f \to f_u ^-$, 
the molecule the average number $<n>$ of open base pairs 
undergoes a continuous power-law
divergence.  From the probability to have $n$ open base pairs, 
$P(n)=  {g(f)}/({k_B T})\; \exp[- (n\, g(f))/({k_BT})]$,
one obtains
\begin{equation}
<n>=\frac{ k_B T}{g(f)} \simeq \frac 1 {f_u -f}
\end{equation}

\begin{figure}[t] 
\begin{center}
\psfig{figure=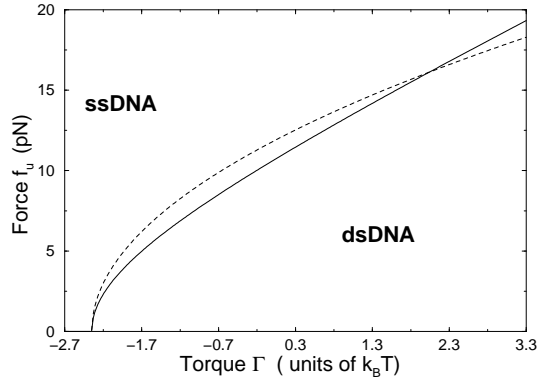,height=7cm,angle=-90} \break 
\end{center}
\caption[]{DNA unzipping phase diagram as a function of torque
$\Gamma$ in units of $k_B T$ and force $f$ in pN, from [54]. 
The solid line shows
the results for the FJCL model of ssDNA elasticity, while the dashed
line shows the result within the Gaussian approximation. At zero torque,
the unzipping force is $f_u \simeq 12$ pN; positive torque increase $f_u$,
negative torques reduce $f_u$ until it vanishes at $\Gamma= -2.4 k_B T$. }
\label{phasediagfig}
\end{figure}

When the DNA molecule is subjected to a torque $\Gamma$,
a torque-angle work contribution occurs in the unzipping free energy,
\begin{equation}
\label{fct}
g (f,\Gamma)= 2 {\cal F}_{ss}(f) +g_0 - \Theta_0\Gamma \qquad .
\end{equation}
$\Theta_0=2\pi/10.4$ is the change in twist angle during conversion of
dsDNA to separated strands\cite{Coc01}. 
The phase diagram for the unzipping transition in the force, torque
plane is shown in Fig.~\ref{phasediagfig}.

\begin{figure}[t] 
\begin{center}
\psfig{figure=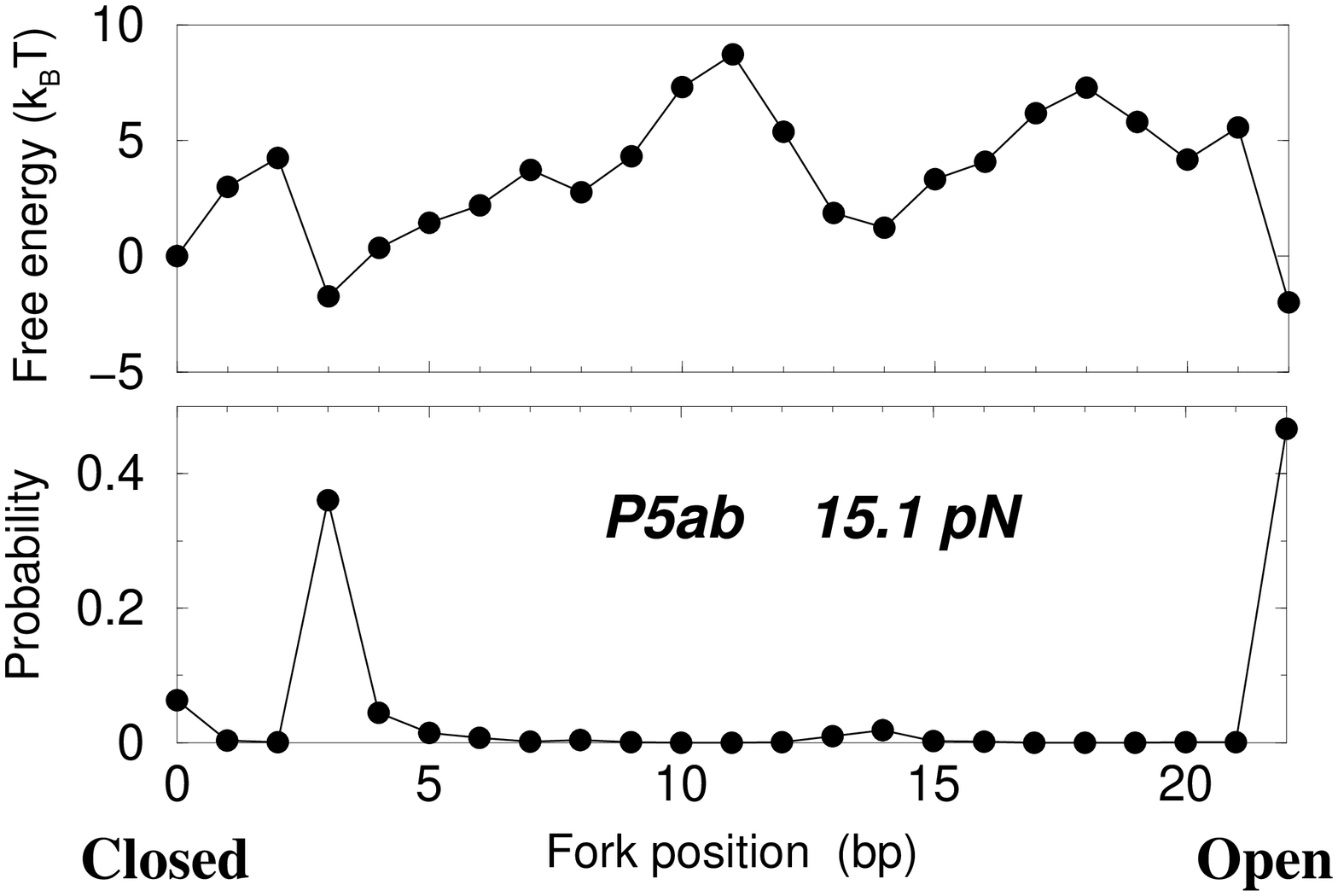,height=6cm,angle=-90} \break
\psfig{figure=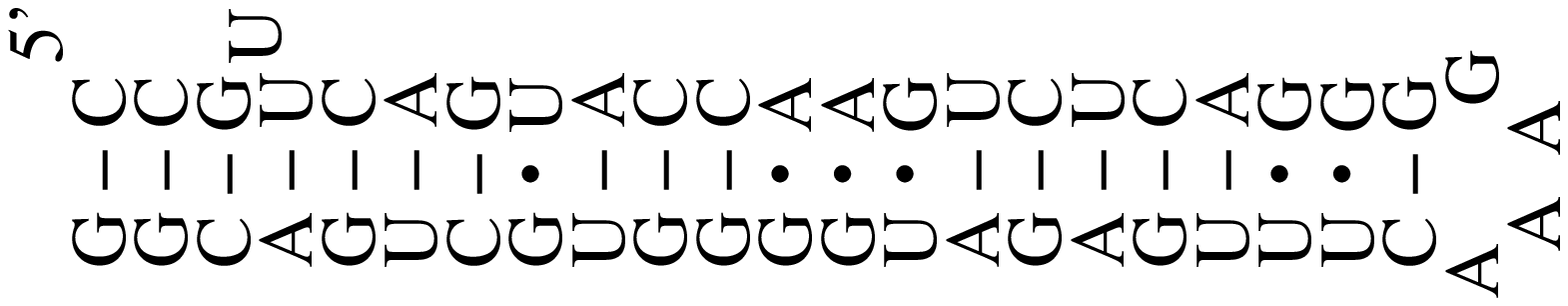,height=5.7cm,angle=90} \break 
\psfig{figure=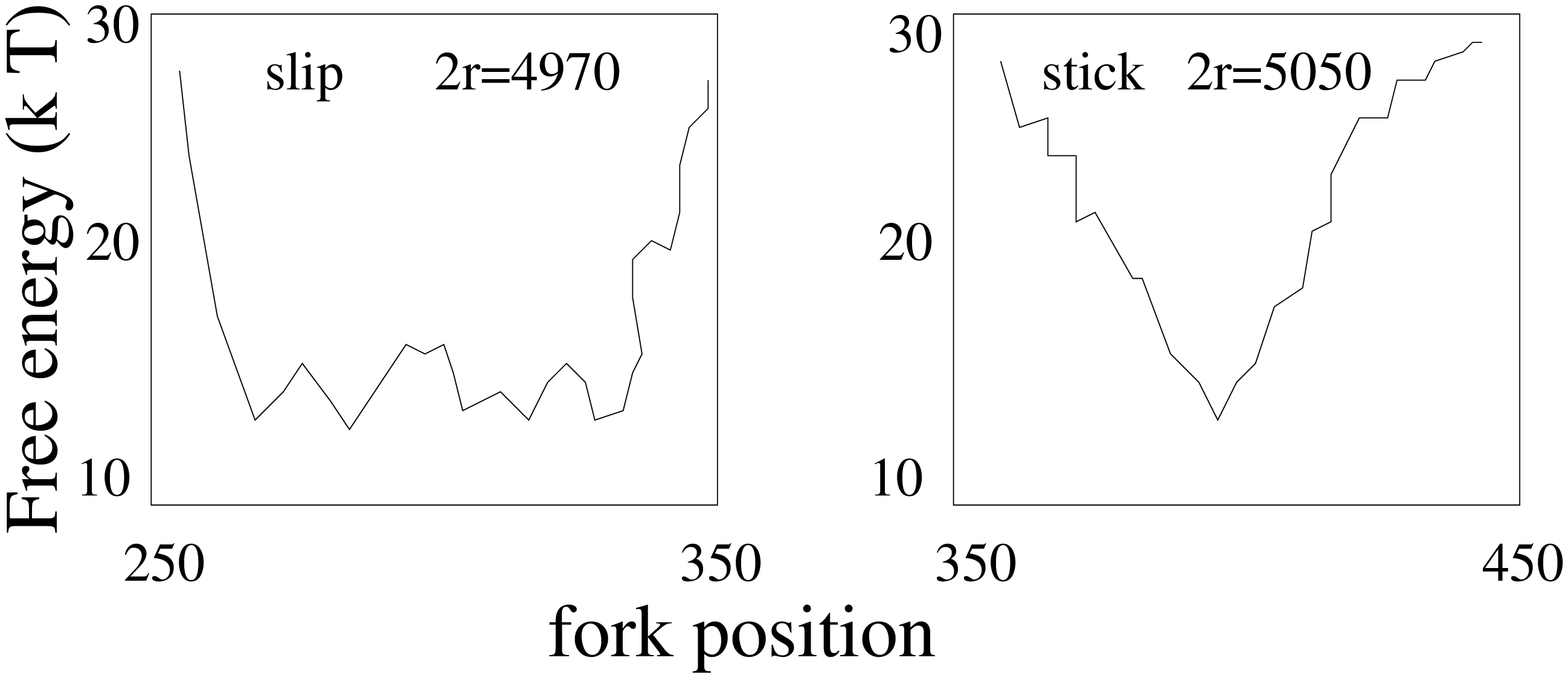,height=5cm,angle=0}
\caption[]{Top: Free energy (in $k_B T$) and probability
 distribution of the opening fork at the critical force
$f=15$ pN for the P5ab molecule (middle) from [59].
Bottom: Schematic representation of the free energy landscape
for a displacement of 4970 nm (left, corresponding to slip phase in the 
force) and of 5050 nm (right, corresponding to the stick phase in the 
force) during the unzipping of a $\lambda$-DNA (from [58]). 
\label{p5abfig}}
\end{center}
\end{figure}

Along heterogeneous sequences, the free energy to open the first $n$ base
pairs,  $G(f,n)$ (\ref{unz1}), can be calculated using the sequence dependent
pairing free energy $g_{ds}(i)$ (e.g. from the Mfold server \cite{Zuk00}).
In Fig.~\ref{p5abfig}, we show $G(f,n)$  for the RNA molecule called
P5ab, mechanically  unzipped by Liphardt {\em et al.} \cite{Lip01} with
a force maintained fixed at the extremity through a feedback mechanism.
The critical force is defined by the condition
that the closed  and open state have equal minimal free energies.
Contrary to the homogeneous case,
the free energy landscape at the critical force is not flat.
It is characterized by high energy barriers $G^*\approx 10 k_B T$.
The probability to have $n$ open base pairs is essentially zero
if $n$ differs from the open and the closed configuration
(Fig~\ref{p5abfig}). Indeed, experiments show \cite{Lip01} that
at the critical force the molecule essentially hops between open and 
closed configurations.        

Lubensky and Nelson \cite{Lub00,Lub02} have shown that
the critical behavior around $f_u$ changes for  large random sequence
with respect to  homogeneous sequences. Instead of the divergence 
$1/(f_u-f)$ for the averaged number $<n>$ of base pairs, a stronger
singularity $<n> \approx 1/(f_u-f)^2$ appears.

\subsubsection{Fixed distance between extremities}
\label{refutile}
If the ssDNA ends are held apart at some 
distance $2r$ (Fig. \ref{unzfig}), some average number of bases $n$ will open.
In the  ideal case of
a rigid opening device, the free energy cost to open $n$ base pairs
is a sum of chain stretching and denaturation contributions,
\begin{equation}
\label{fixdis}
H_{d} (r,n)=
 W_{ss}(2r,2n) -\sum_{i=1}^{n} g_{ds}(i)
\end{equation} 
$W_{ss}(2r,2n)$ is the work done by the force to stretch  $2n$ base pairs
of ssDNA at a distance $2r$.  For simplicity we consider
the GP free energy (\ref{spring}),
considering as in the previous section the effective Kuhn length $b=7 \AA$~,
from the interpolation formula (\ref{mark2}) 
\cite{Tho95}, or from numerical inversion of the FJC extension
vs. force (\ref{fjc2})~\cite{Boc98}.
Thus, unzipping at fixed extension can bedescribed 
using 
\begin{equation}\label{gcea}
W^{GP}_{ss}(2r,2n)= 3 k_B T \; \frac{r^2} {n\; b^2} 
\end{equation}

The most probable value of the number of opened base pairs $\bar n$ is 
obtained by minimization of the free energy (\ref{fixdis},\ref{gcea})
with respect to $n$. 
The number of unzipped based pairs is found to scale linearly with
the distance, $\bar n (r)= r/d_u $
where $d_u=\sqrt{g_0/3 k_B T } d =5 \AA$~ 
is the projection of the monomer length along the force direction.
The resulting free energy reads 
\begin{equation}
{\cal F}(r)=2\,\sqrt {3 k_B T g_0}\;\, r/b= 2\, \bar n (r)\; g_0
\end{equation}
The tension $\bar f$ in the chain is simply the derivative of ${\cal F}$  
with respect to $2r$,
$\bar{f}= \sqrt { 3 k_B T g_0}/b = 12$ pN.
This simple calculation shows that as unzipping proceeds quasi-statically,
the ssDNA tension is a constant, just $f_u$.
Note that the excess free energy per unpaired base for fixed extension
is double the free energy of denaturation because the
work done extending the ssDNAs adds to the work done when opening the 
molecule.   This indicates a strategy to determine $g_{ds}$ 
unambiguously from unzipping force data.

The analysis of the fluctuations around the minimum free energy gives 
$\left< f  -f_u \right> = O(1/r ^2) $. Note that in the 
constant--force ensemble result, $f -f_u \approx  1/<r>$ \cite{Lub02}. 
The fixed-distance and fixed-force ensembles are equivalent only in the 
thermodynamic limit $r \to \infty$.

The unzipping force at small distance between extremities is
sensitive to the detailed structure of the pairing potential 
as a function of the interbase distance. The semimicroscopic model
introduced in \cite{Coc01,Coc02} predicts a force barrier of $\simeq
300$~pN  at a  distance $2r \simeq 21 \AA$~,
at which the hydrogen bond are broken but the bases are still stacked in
the double helix configuration.  This force barrier might be directly
observable in an AFM experiment.

To take into account the experimental apparatus, 
Bockelmann {\em et al.} have included in their theory
the effects of the two  dsDNA linker arms
of $N_{ds}$ base pairs and extension $r_{ds}$),
 and the cantilever stiffness \cite{Boc98,Boc02}. 
The free energy (\ref{fixdis}) is then
\begin{equation}
H_d^c (r,r_{ss},r_{ds},n)=
W_{ss}(2 r_{ss}, 2n)+W_{ds}(r_{ds},N_{ds}) -\sum_{i=1}^{n} g_{ds}(i) -
 k_{lever} (r-2r_{ss}-r_{ds}) \ .
\end{equation}
The partition function is obtained by summing over all possibles value of 
$n,r_{ss},r_{ds}$. The free energy $g_{ds}(i)$ can be computed
using e.g. the Mfold program for base-pair interactions \cite{Zuk00}.
The ssDNA, dsDNA and the lever can be considered as three springs
in series, with an  effective stiffness is 
$k_{tot}^{-1}=k_{ss}^{-1} + k_{ds}^{-1} +k_{lever}^{-1}$.
The spring constant of dsDNA, 
$k_{ds}  \simeq 0.03$~ pN/nm for a 
dsDNA total length of $15000$ bases in the experiment
of Bockelmann {\em et al.},   is obtained from the
derivative of $f^{WLC}$ (including the Young modulus)
calculated  at a force of 12 pN. The ssDNA stiffness is
$k_{ss} \simeq 6 k_{B T} /(b^{2} n) \simeq 50/n$~pN/nm, and
the cantilever stiffness equals $k_{lever}=0.25$~ pN/nm.
For less than $\approx 1500$ unzipped base pairs, $k_{tot}$ essentially
reduces to the dsDNA linker stiffness. When more bases are unzipped,
the dominant contribution comes from the ssDNA stiffness.

The DNA sequence dependence results in a complicated free energy landscape
that generates a 'stick-slip' variation of the force during unzipping
\cite{Boc02}.
The stick phase corresponds to the presence of one deep
minimum, and the slip phase to a flat free energy landscape 
see Fig~6. The analytical description of Bockelmann 
{\em et al.} predicts that the
 height $h$ of the potential barrier increases as $h \approx \delta^2$
with the fluctuations $\delta$ of the pairing free energy,  
and decreases  $h \approx k_{tot}^{-1}$ with the effective stiffness.

\subsection{Kinetics of unzipping}

The kinetics of unzipping at short length scales
is affected by the presence of barriers
with various physical origins, which makes it an activated process.

\subsubsection{Homogeneous sequences and unzipping initiation barriers}

\begin{figure}
\begin{center}
\psfig{figure=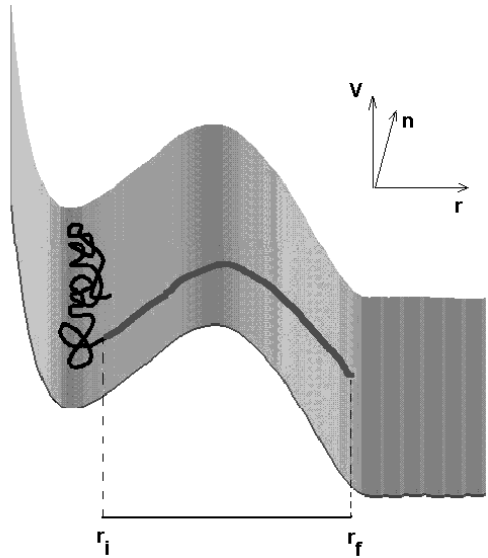,height=7cm,angle=-90} 
\end{center}
\vskip .5cm
\caption{Schematic representation of the saddle-point polymer
conformation $r^*(n)$ which crosses the barrier of $V(r)$;
$r_i$ and $r_f$ are the radii of the edges of the nucleation bubble.}
\label{rub}
\end{figure}

Unzipping of homogeneous DNA requires the crossing of a barrier
whose physical origin is the following.  We imagine $r$, the half
distance between the two bases of one pair to be a reaction coordinate
indicating whether the pair is bonded ($r\simeq 10 \AA$) or open
($r>11-12 \AA$~ due to the very short range of H-bond). 
The effective free energy $V(r)$
of the base pair as a function of $r$ is shown in Fig.~\ref{rub}.
It is low for both small (pairing energy) and large (entropy gain)
values of $r$, and exhibits a maximum around $r\simeq 10.5 \AA$~ where
the H-bond is broken but bases are still stacked in the double helix 
conformation and are not free to move \cite{Coc01}. 
The set of half distances $r(n)$ between the two strands defines an
abstract polymer (Fig.~\ref{rub}). At low enough forces this polymer
is confined to the potential well (closed state). When a force $f$ larger
than $f_u$ is applied at one extremity, the polymer escapes from
the well (unzipping). As in a first order phase transition, nucleation
theory can be employed to understand the opening kinetics\cite{Lan67}. 
  
The kinetics are slowed because of the activated
crossing of the free energy barrier.
A saddle-point calculation provides the optimal configuration of the
polymer for crossing the barrier\cite{Coc01}. This configuration is made of a 
transition `bubble' of a few $\simeq 4$ bases long, 
and free energy cost $G^*(f)$.
The time of unzipping initiation grows exponentially with
the activation free energy $G^*(f)$, $t(f)=t_0 \exp (G^*/k_BT)$.
The elementary time $t_0$ corresponds to the time necessary for the
polymer to escape from the saddle-point configuration along the unstable 
direction in the free energy landscape\cite{Lan67}.
When the applied force is smaller than $f_u$, the molecule may 
still unzip. The opening time, which still depends on the barrier
free energy, is now exponentially large in the equilibrium 
free energy $N g(f)$ where $N$ is the number of base pairs.

The determination of the opening time $t(f)$ at fixed force provides
in turn the distribution of unzipping forces when the molecule is
loaded with a fixed rate\cite{Eva92}. The most probable unzipping 
force exhibits a rich pattern
depending on the loading rate and on the length of the 
sequence\cite{Coc01,Coc02}.

\subsubsection{Sequence-dependent barriers}

The above mechanism is mostly relevant in the kinetics of initiation
of opening of the double helix.   During unzipping of a long
double helix, the sequence dependent free energy
landscape of Fig~\ref{p5abfig} with barriers of $\approx 10 k_BT$ is
responsible for a slow hopping dynamics between the open and closed
states, i.e. the slip-stick behavior observed by Bockelmann et al
\cite{Ess97,Boc02}.  

In experiments by Liphardt et al\cite{Lip01}, small helix-loop RNA structures
(essentially short regions of double-helical RNA terminated with a short
loop) were held in such a way that equilibrium fluctuation between
open and closed states occurred.  The timescale observed was
close to 1 sec \cite{Lip01}, remarkably long for a few-nm-long molecule.
In \cite{Coc03b} a dynamical model was introduced for the motion of
the `fork' separating the base paired and opened regions of the
molecule, allowing computation of the opening and closing rate as a
function of the force. 
The model, which describes the experimental data well,
is based on the following elementary
rates of opening and closing base pair $n$
\begin{equation}
r_0(n)= r\; e^{-g_0(n)/k_B T}, \;\; r_c(f,n)= r\; e^{-2 {\cal F}_{ss,}
(f,n)}
\end{equation}
The opening rate is taken to depend only on the pairing free energy since 
the short range hydrogen bond is broken before the force-length work
over the longer $\approx 0.7$ nm distance can be done.   Conversely,
to close the base pair, work must be first done against the
applied force, and so the closing rate is taken to be depend only on
the force.  
The separation of length scales of the range of base
pair interaction and base pair extension after unzipping is thus
used to justify placing most of the force-dependence in the zipping rate, 
with most of the interaction dependence in the unzipping rate.
More sophisticated rate models will require further experiments to
determine their form.

\section{Conclusion}

We have presented a very brief overview of the theory used to think
about single-molecule nucleic acid micromanipulation experiments.  
The field of single-molecule experiments is evolving so rapidly at present 
that we have been forced to omit many exciting topics.
Here we present a few general comments about
what has been learned and suggest some directions that might be particularly
interesting for study in the near future.

A feature common to all the studies described above, and to the theory 
of other types of single-biomolecule
experiments, is the central role of statistical mechanics.
The interaction of this field with statistical mechanics is fundamental:
the understanding (in some cases, even the primary data analysis)
of single-molecule DNA experiments requires statistical
mechanics.  Additionally, previously unimagined 
statistical-mechanical problems are resulting from
the huge range of experimental possibilities for DNA and DNA-protein 
micromechanical experiments.

The first phase of single-DNA experiments involved basic characterization
of dsDNA and ssDNA, and from the theoretical side involved development
and solution of statistical-mechanical theories for the molecules subjected
to stresses.   The studies reviewed above essentially fall into this
first class, and are characterized by a degree of 
quantitative success, thanks both to the efforts of experimentalists
and theorists, which is unprecedented in soft condensed matter physics.

The second phase, which we are in at present, involves
the study of modifications of the basic molecules, e.g. by unzipping,
or by action of proteins acting on DNAs under 
mechanical control.   The statistical mechanics of this second class
of problems is less well developed, and includes problems far from
thermal equilibrium such as DNAs acted on by processive, ATP-powered
motor-like enzymes.   The diversity of challenging and new statistical
physics problems in this second class is mind-boggling.
The second phase is also forcing theorists to confront the
information content of nucleic acids since sequence plays an
essential role in targeted nucleic acid-protein interactions.

The third class of problems involves applications of the lessons
learned, to the description of cell machinery in vivo, or at least
under in-vivo-like conditions.   Such experiments are in their
infancy, and extension of theoretical physics into this domain 
is still in a dark age.  However, we can look forward
to a time when we understand processes such as cell division and
growth, gene regulation, and other cell biological functions in
statistical-mechanistic terms.  The lessons being learned now
about the importance of statistical mechanical ideas to 
biochemical-micromechanical experiments on nucleic acids will thus become
an important component of future quantitative understanding 
of cell biology.

{\bf Acknowledgements}
We thank A. Sarkar, V. Croquette and D. Bensimon for their helpful comments
on this manuscript.  This work was in part supported by
the NSF(USA; JM and SC: DMR-9734178, RM: DMR-9808595), 
the Research Corporation (JM), the A. della Riccia Foundation (SC),
and by the Focused Giving Program of Johnson and Johnson (JM).

\end{document}